\documentclass[11pt]{article}
\providecommand{\keywords}[1]{\textbf{\textit{Keywords:---}} #1}

\usepackage{authblk}
\usepackage{latexsym,amsmath,amsfonts,amsthm,amssymb,bbm,fdsymbol}
\usepackage{epsfig,graphics,color,calc,graphicx}
\usepackage{color}

\usepackage{color}

\numberwithin{equation}{section}
\theoremstyle{plain}
\newtheorem{thm}{Theorem}
\newtheorem{lemma}[thm]{Lemma}
\newtheorem{corollary}[thm]{Corollary}

\theoremstyle{definition}

\newcommand{\mmmintone}[1]{{\dis{\int\kern -.36cm
-}}_{\kern-.21cm\substack{#1}}\;\;}
\newcommand{\mmmintwo}[2]{{\dis{\int\kern -.43cm
-}}_{\kern-.21cm\substack{#1}}^{\substack{#2}}\;\;}
\newcommand{\submint}{{\scriptstyle{\int\kern -.66em -}}}
\newcommand{\submintone}[1]{{\scriptstyle{\int\kern -.66em
-}}_{\scriptscriptstyle{\kern-.21em\substack{#1}}}}
\newcommand{\fracmint}{{\textstyle{\int\kern -.88em -}}}
\newcommand{\fracmintone}[1]{{\textstyle{\int\kern -.88em
-}}_{\scriptscriptstyle{\kern-.21em\substack{#1}}}\;}

\newcommand{\dis}{\displaystyle}

\newcommand{\nn}{\nonumber}

\newcommand{\la}{\lambda}
\newcommand{\La}{\Lambda}

\newcommand{\eps}{\epsilon}
\newcommand{\ga}{\gamma}

\newcommand{\si}{\sigma}

\newcommand{\nada}[1]{}

\def\square{\ifmmode\sqr\else{$\sqr$}\fi}
\def\sqr{\vcenter{
         \hrule height.1mm
         \hbox{\vrule width.1mm height2.2mm\kern2.18mm\vrule width.1mm}
         \hrule height.1mm}}                  

\title{Microscopic models for uphill diffusion}
\author{M. Colangeli \footnote{Universit\`{a} degli Studi dell'Aquila, Via Vetoio, 67100 L'Aquila, Italy.\\ E-mail: matteo.colangeli1@univaq.it}, A. De Masi\footnote{Universit\`{a} degli Studi dell'Aquila, Via Vetoio, 67100 L'Aquila, Italy.\\ E-mail: anna.demasi@univaq.it}, E. Presutti \footnote{Gran Sasso Science Institute, Viale F. Crispi 7, 00167 L'Aquila, Italy.\\ E-mail: errico.presutti@gmail.com}}
\date{\today}

\begin{document}

\maketitle

\begin{abstract}
\noindent
We study a system of particles which jump on the sites of the interval $[1,L]$ of $\mathbb Z$.  The density at the boundaries is kept fixed to simulate the action of mass reservoirs.  The evolution depends on two parameters $\la'\ge 0$ and $\la''\ge 0$ which are the strength of an external potential and respectively of an attractive potential among the particles.  When
$\la'=\la''= 0$ the system behaves diffusively and the density profile of the final stationary state is linear, Fick's law is satisfied.  When $\la'> 0$ and $\la''= 0$ the system models the diffusion of carbon in the presence of silicon as in the Darken experiment: the final state of the system is in qualitative agreement with the experimental one and uphill diffusion is present at the weld.   Finally if  $\la'=0$ and $\la''>0$ is suitably large, the system simulates a vapor-liquid phase transition and we have a  surprising phenomenon.  Namely when the densities in the reservoirs correspond respectively to metastable vapor and metastable liquid we find a final stationary current which goes uphill from the reservoir with smaller density (vapor) to that with larger density (liquid).  Our results are mainly numerical, we have convincing theoretical explanations  yet we miss a complete mathematical proof.

 \end{abstract}

\noindent

\vskip2cm

\section {Introduction}
\label{sec:intro}

Uphill diffusion is a phenomenon which appears when the current flows  {\em along} the gradient in contrast with the Fick law which states that the current is proportional to {\em minus} the gradient.  We are considering the case of mass diffusion so that the current is the mass flux and the gradient is the gradient of the mass density.

There are mainly two cases where uphill diffusion appears, the first one is when the system is a mixture of two or more components (or one component but several conserved quantities), the second one when the system undergoes a phase separation.  The macroscopic explanation in the first case is that the current of say component 1 has a contribution $j_1$ proportional to minus the gradient of its density (in agreement with the Fick law) but also contributions coming from the gradients of the other components which may be larger than $j_1$ and with the opposite sign.  The basic reference is an old paper by Darken, \cite{darken}, where he gives experimental evidence of the phenomenon \cite{Tsuc,Vitag}, see also \cite{Karg,RK} for an updated survey and Refs. \cite{Alvarez,Diebel,Erleb,Frink,Lauer,Sato} for related recent results. We will present here a simple particle model which reproduces qualitatively the main features of the Darken experiments.

Uphill diffusion appears also in one component systems at phase transition, say a vapor-liquid transition.  In such a case there is a density  interval $(\rho',\rho'')$ so that if the density$\rho$ is $\le \rho'$ the system is in its vapor phase while if $\rho \ge \rho''$ then it is in its liquid phase. If we put a mass $\rho|\La|$ of fluid in a region $\La$ with $\rho \in(\rho',\rho'')$ we observe a mass flux  which gives rise to  a non homogeneous final density profile with vapor at density $\rho'$ in a subregion $\Delta$ of $\La$ while in the complement the phase is liquid at density $\rho''$.  Thus during the phase separation mass has flown from the lower density in $\Delta$ to the larger density in $\La \setminus \Delta$: it has gone uphill.

We  call this phenomenon a ``transient uphill diffusion'' as in the end there is no current and we  distinguish it from a
``steady uphill diffusion''. The latter arises when a single component fluid in contact with a left and a right mass reservoir at density $\rho_-<\rho_+$ reaches a stationary state with positive current, namely mass flows from the reservoir at lower density to the one at larger density.  We have observed this phenomenon in computer simulations of a particle system which models a vapor-liquid phase transition. The
system is put in contact with mass reservoirs which keep fixed the densities at the boundaries. We have seen that if the left reservoir fixes a density $\rho_-$ in the metastable vapor phase while the right reservoir density is $\rho_+$ in the metastable liquid phase  then the system reaches a stationary state where the current is positive, namely flows from left to right, i.e.\ mass goes from the reservoir at small density to that with larger density. Instead
if the reservoirs densities are in the vapor and liquid stable phases (i.e.\ $\rho_-< \rho'$, $\rho_+> \rho''$), then the current is negative and goes downhill.

The final stationary state when the current is positive could be either one where in most of the space the fluid is liquid with a small region close to the left boundary where there is a sharp transition from vapor to liquid or symmetrically one where in  most of the space the fluid is vapor with a small region close to the right boundary where there is a sharp transition from liquid to vapor.  In both cases the density profile is decreasing except in the transition region at either one of the boundaries, thus the steady current goes downhill in most of the space and uphill at the transition.  The final stationary state is determined by the initial conditions and by random fluctuations.

We are not aware that such a ``steady uphill diffusion'' has been observed earlier and it certainly deserves to investigate whether analogous phenomena are present in more general systems.

In Section \ref{sec:2} we consider a particle model which describes normal diffusion; by adding a suitable potential we obtain in  Section \ref{sec:3} a system which simulates the Darken experiment with carbon diffusing in the
presence of silicon. In Section \ref{sec:4} we modify the model of Section \ref{sec:2} by adding an attractive force among the particles which gives rise to uphill diffusion in the presence of phase transition.  In Appendix \ref{appA} we prove a theorem stated in Section \ref{sec:3}.

\bigskip

\setcounter{equation}{0}

\section{A  microscopic model for diffusion}
\label{sec:2}

We want to describe fluids where the evolution is diffusive and convection is negligible.  These are the main features that our models should try to
catch.

\begin{itemize}

\item  Particles undergo very frequent collisions, their velocities change rapidly and erratically  so that their motion looks diffusive, convection being absent.

\item  There is a strong repulsive force when particles are too close to each other which makes the density bounded.

\item  The system is confined in a cylindrical vessel with a horizontal axis. The two extremal faces are in contact with mass reservoirs which keep the mass density at the boundaries fixed.  We suppose a planar symmetry in the vertical planes orthogonal to the axis of the cylinder.

\end{itemize}

\vskip.5cm

\noindent
We will first describe our model designed for implementation on a computer, then show some computer simulations and finally  discuss how well it catches the above physical requests.

The model is one dimensional, space and time are discrete. Particles are confined in the interval $\{1,...,L\}$: $L$, a positive integer, is the spatial size of the system. Particles have only velocities equal to $+1$ and $-1$. Time is discrete: $t=0,1,2,\dots$ so that at each time step particles move from one site to the next one (right or left according to their velocity), we will say later what happens at the boundaries.  There is an exclusion rule which prevents two particles with same velocity to stay on a same site, thus the local density is necessarily $\le 2$. Before moving particles change randomly their velocities, however the exclusion rule prevents changes at sites where there are two particles (as they must have opposite velocities).  The precise algorithm used to update the particles configurations is as follows.

Particles configurations are described by sequences $\eta= \{\eta(x,v), x\in [1,L],v\in\{-1,1\}\}$ with $\eta(x,v)\in \{0,1\}$  the occupation variable at the phase space point $(x,v)$.  We denote by $\eta(x)=\eta(x,-1)+\eta(x,1)$ the total
occupation at $x$ and  add a suffix $t$ when the occupation variables are computed at time $t$, $t=0,1,2\dots$.
The unit time step updating is obtained as the result of three successive
operations starting from a configuration $\eta$ and ending with a configuration $\eta'''$, we denote by   $\eta'$ and $\eta''$ the configurations at the intermediate steps.

\begin{enumerate}
\item {\em velocity flip}.  At all sites $x\in [1,L]$ where there is only
one particle we update its velocity  to become $+1$ with probability $\frac 12
$ and $-1$ with same probability $\frac 12$ (such velocity flips are independent of each other).
At all other sites the occupation numbers are left unchanged. We denote by $\eta'$ the occupation numbers after the  velocity flip updating.

\item {\em advection}. After deleting the particles at $(1,-1)$ and $(L,1)$
(if present) we  let
each one of the remaining particles move by one lattice step in the direction of their velocity.
  We denote by $\eta''$ the occupation numbers after this advection step.

\item {\em boundaries updating}.  Let $\rho_{\pm} \in [0,1]$ and call
$2\rho_{\pm}$ the density of the right, respectively left reservoir.  Then
with probability $\rho_+$ we put a particle at $(L,-1)$ and with probability  $1-\rho_+$ we leave  $(L,-1)$ empty.  We do independently the same operations at $(1,1)$ but with $\rho_-$ instead of $\rho_+$.  What we get is the final configuration $\eta'''$.

\end{enumerate}

\medskip

\noindent
Let us next see how the model behaves.  We have run several computer simulations, we report below some of them.  It is convenient here and in the sequel to change variables writing
   \begin{equation}
        \label{2.1}
\si_t(x)=\eta_t(x) -1,\quad m_{\pm}= 2\rho_{\pm} -1
    \end{equation}
thus $\si_t(x)\in \{-1,0,1\}$ and $m_{\pm} \in [-1,1]$.  The above change of variables, which simplifies some formulas below,  has also a physical meaning
in terms of magnetic systems with $\si_t(x)$  a spin, we  refer to \cite{CDP_jsp} for details.  We fix an initial datum where the variables $\eta(x,v)$
are independent and take values $0,1$ with same probability.  For any choice of the initial datum we run the above algorithm for a time $t_0+T$ and measure for each $x \in [1,L]$ the time average
   \begin{equation}
        \label{2.2}
\si^{t_0,T}(x)=\frac 1T\sum_{t=t_0+1}^{t_0+T} \si_t(x)
    \end{equation}
The current at time $t$ from the system to the right reservoir is
   \begin{equation}
        \label{2.3}
j_{+}(t) =  \eta_t(L,1) - \eta_{t+1}(L,-1)
    \end{equation}
which counts as positive the particles which leave the system from the right and as negative those which enter from the right.  Analogously  the current at time $t$  from the the left reservoir to the system is
   \begin{equation}
        \label{2.4}
j_{-}(t) =  \eta_{t+1}(1,1)-\eta_t(1,-1)
    \end{equation}

\begin{figure}
\centering
\includegraphics[width=0.6\textwidth]{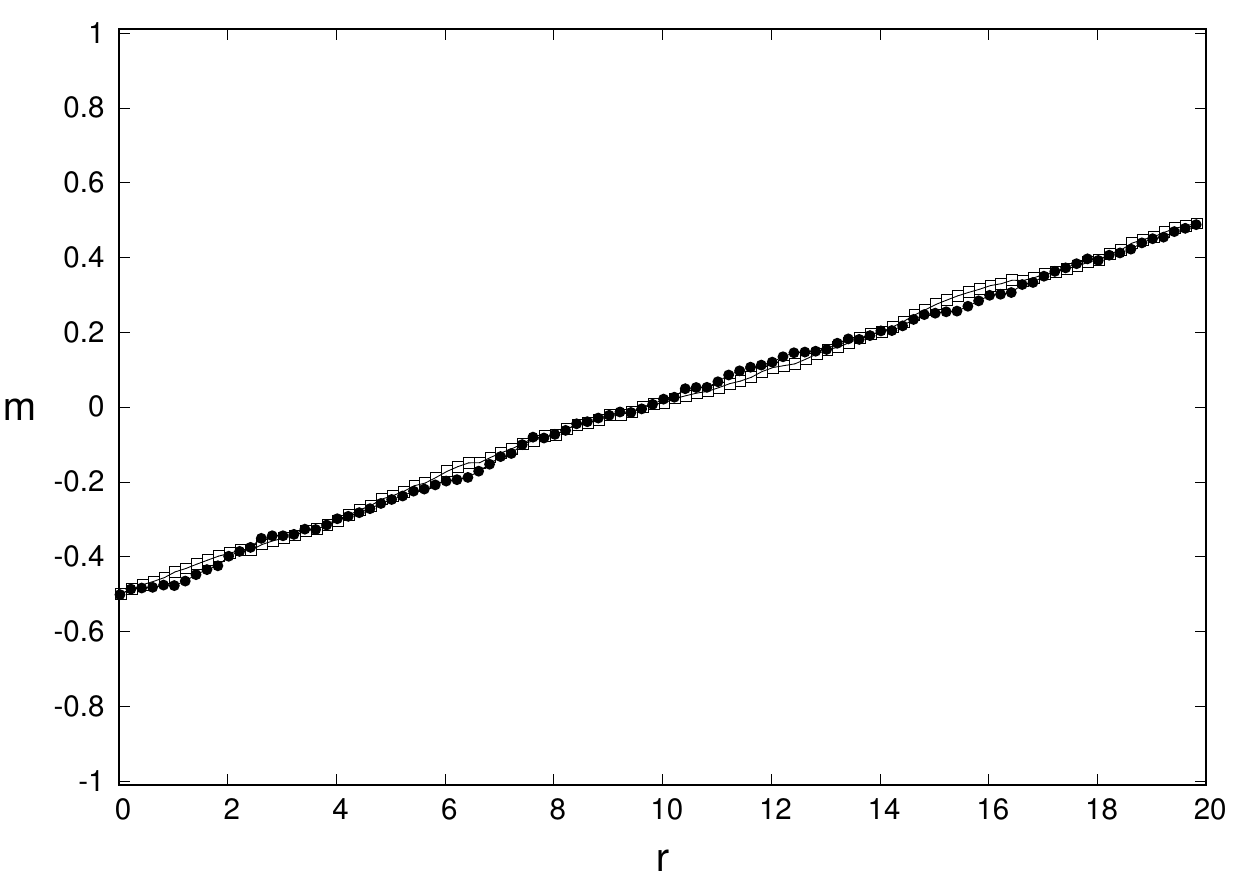}
\caption{Magnetization profile corresponding to a purely diffusive behavior, with $m_{+}=-m_{-}=0.5$.
Open and solid symbols are the Monte Carlo prediction
for, respectively, $L=600$ ($\smallsquare$) and $L=1200$ ($\smallblackcircle$).}
\label{fig:fig1}
\end{figure}

We have also measured the averaged currents
   \begin{equation}
        \label{2.5}
j_{\pm}^{t_0,T}=\frac 1 T\sum_{t=t_0+1}^{t_0+T} j_{\pm}(t)
    \end{equation}
The simulations we report here are done with $L= 600$ and $L=1200$, $t_0=10^9$, $T=10^5$,  $m_+=0.5$,
and $m_-=-0.5$ in both cases. In Fig. \ref{fig:fig1} we have plotted $m_L(r), r \in[L^{-1},1]$ by setting
$m_L(r) = \si^{t_0,T}(Lr)$ when $L=600$ and $L=1200$, the difference between the two profiles is negligible.  The currents $j_{\pm}^{t_0,T}$ are essentially equal to each other and
   \begin{equation}
        \label{2.6}
j_{\pm}^{t_0,T}\approx \frac 12 \frac{m_+-m_-}{L},\qquad L=600,\,L=1200
    \end{equation}
In conclusion the simulations show that the system obeys the Fick law with constant diffusion coefficient equal to 1.  For $L$ large the stationary profile $m_L(r)$ is linear connecting $m_-$ to $m_+$.

\vskip.5cm

Let us finally discuss how well the particle model catches the physical requests stated at the beginning of the section.  As shown from the simulations it indeed describes a diffusive fluid.  The flip velocity updating however is not realistic, the collisions in a real fluid are not as simple and there are correlations between successive collisions, statistical independence that we assume here is the main issue in the derivation of the Boltzmann and other kinetic equations.  The assumption that the speed is 1 is also unrealistic but it is the easiest way to achieve a description on the lattice which is more easily implementable on the computer. In the model the local density $\eta_t(x)$ is always $\le 2$, in real systems a bound on the density comes from strong repulsive forces at short distances, as in Lennard-Jones, in our model it is simply achieved by forbidding velocity flips when two particles are on the same site.  Clearly it is not the true reason yet it does the job.  The restriction to one dimension reflects the assumption of planar symmetry on vertical planes and it is therefore quite acceptable.  The action of the reservoirs updating is to keep the average density at $(L,-1)$ equal to $\rho_+$ and at $(1,1)$ equal to $\rho_-$.  From the simulations we find that in average $\eta_t(x,v)\approx \eta_t(x,-v)$ so that the action of the reservoirs is to fix in the average the density at $1$ and $L$ equal to $2\rho_{\mp}$, i.e.\ the density of the reservoirs.

\bigskip
\setcounter{equation}{0}

\section{The Darken experiment}
\label{sec:3}

In \cite{darken} Darken reports of experiments which show uphill diffusion of carbon, we refer in particular to the case of Fig. 2 in \cite{darken} where carbon diffuses in a welded specimen where the silicon content is concentrated on the left of the weld (and negligible on the right).  We refer to \cite{darken} for the details of the experiment.  We model the carbon atoms using the particle model of the previous section but we need to modify the updating rules to take into account the presence of silicon.  As stated in  \cite{darken} we may neglect the diffusion of silicon so that we suppose that the stationary silicon density is equal to 1 (in appropriate density units) to the left of the weld and to 0 afterwards, i.e.\
$\rho_{\rm si}(x)=\mathbf 1_{x \le \frac L2}$. Carbon does not like to stay where the silicon is, hence the carbon will feel  a positive potential $U_\ga(x)$:
   \begin{equation}
        \label{3.1}
U_{\ga}(x) = \la \sum_{y} J_\ga(y,x) \rho_{\rm si}(y), \quad J_\ga(y,x)= \ga J(\ga|x-y|),\quad \la >0
    \end{equation}
where  $\ga^{-1}$ is a positive integer and
   \begin{equation}
        \label{3.2}
J(r) = (1-r) \mathbf 1_{0\le r\le 1}
    \end{equation}
Namely $U_\ga(x)$ is a sum of the contributions
$J_\ga(y,x) \rho_{\rm si}(y)$ exerted from all silicon atoms.  The interaction strength $J_\ga(x,y)$ has range which scales as $\ga^{-1}$, $\ga^{-1}$ is a parameter of the model which on physical grounds should be  much larger than the interatomic distance (which in our model is the distance between two successive sites and thus equal to 1) but also much smaller than the macroscopic size of the specimen, $L$ in our model.


Our choice of $J(r)$ is of course quite arbitrary, it has the advantage to give a simple formula for the force (defined as usual as minus the gradient of the potential):
  \begin{equation}
        \label{3.3}
f_\ga(x) = -\ga^2 \la \Big(\sum_{y = x}^{x+\ga^{-1}}  \mathbf 1_{y \le \frac L2}
- \sum_{x-\ga^{-1}}^{ x}  \mathbf 1_{y \le \frac L2}\Big)
    \end{equation}
Thus $f_\ga(x)$ is simply equal to $-\ga ^2 \la$ times the difference between the number of silicon particles to the right and left of $x$
in a range $\ga^{-1}$, hence $f_\ga(x)$ is directed to the right
and active only in a neighborhood of the weld:
  \begin{equation}
        \label{3.3.0}
f_\ga(x) = \ga^2 \la \mathbf 1_{|x-\frac L2| \le \ga^{-1}}\times\begin{cases}
x-(\frac L2 -\ga^{-1}),& x \le \frac L2\\
(\frac L2 +\ga^{-1})-x, & x \ge \frac L2 \end{cases}
    \end{equation}
To represent the force in our model we argue
that $\delta v$,
the average velocity change per unit time
due to the force, should be proportional to $\beta f_\ga (x)$, $\beta$ the inverse temperature (as thermal fluctuations dampen the effect of the force).  We can put this
in our model where velocities are only $\pm 1$ by changing the probability of the velocity flips.  We thus modify the updating rules of the previous section only in the flip velocity step: at sites where there is only one particle its velocity
is updated to be $+1$ with probability $\frac 12  + \eps_{x,\ga}$ and equal to $-1$
with probability $\frac 12  - \eps_{x,\ga}$, thus the average velocity after the flip is $2\eps_{x,\ga}$ which has the desired value $\beta f_\ga(x)$  if
  \begin{equation}
        \label{3.4}
\eps_{x,\ga} = \frac 12 \beta f_\ga(x) =  \ga^2 \frac {\beta \la }2 \mathbf 1_{|x-\frac L2| \le \ga^{-1}}\times\begin{cases}
x-(\frac L2 -\ga^{-1}),& x \le \frac L2\\
(\frac L2 +\ga^{-1})-x, & x \ge \frac L2 \end{cases}
    \end{equation}
As a consequence there is  a bias to the right when close to the weld.
(Recall that all the other updating rules are left unchanged).

Since we want to single out the effect of the force due to the silicon we take a homogeneous initial datum where as in Section \ref{sec:2} the variables $\eta(x,v)$ are independent and each one has an average equal to $1/2$.  Then the average density, i.e.\ the average of $\eta(x)$, is equal to 1.  Also the reservoirs have density  $1$, i.e.\ $m_{\pm}=0$.
Having defined the model and the initial datum we can now run the simulations.
We take the size of the system $L$ equal   to 600 or to $1200$, $\ga^{-1}$ equal to 30 or to 60.  We call $r=\ga x$ the space measured in mesoscopic units and
$\ell =\ga L$ the size of the system in mesoscopic units.

\vskip1cm

\begin{figure}[h!]
\centering
\includegraphics[width=0.6\textwidth]{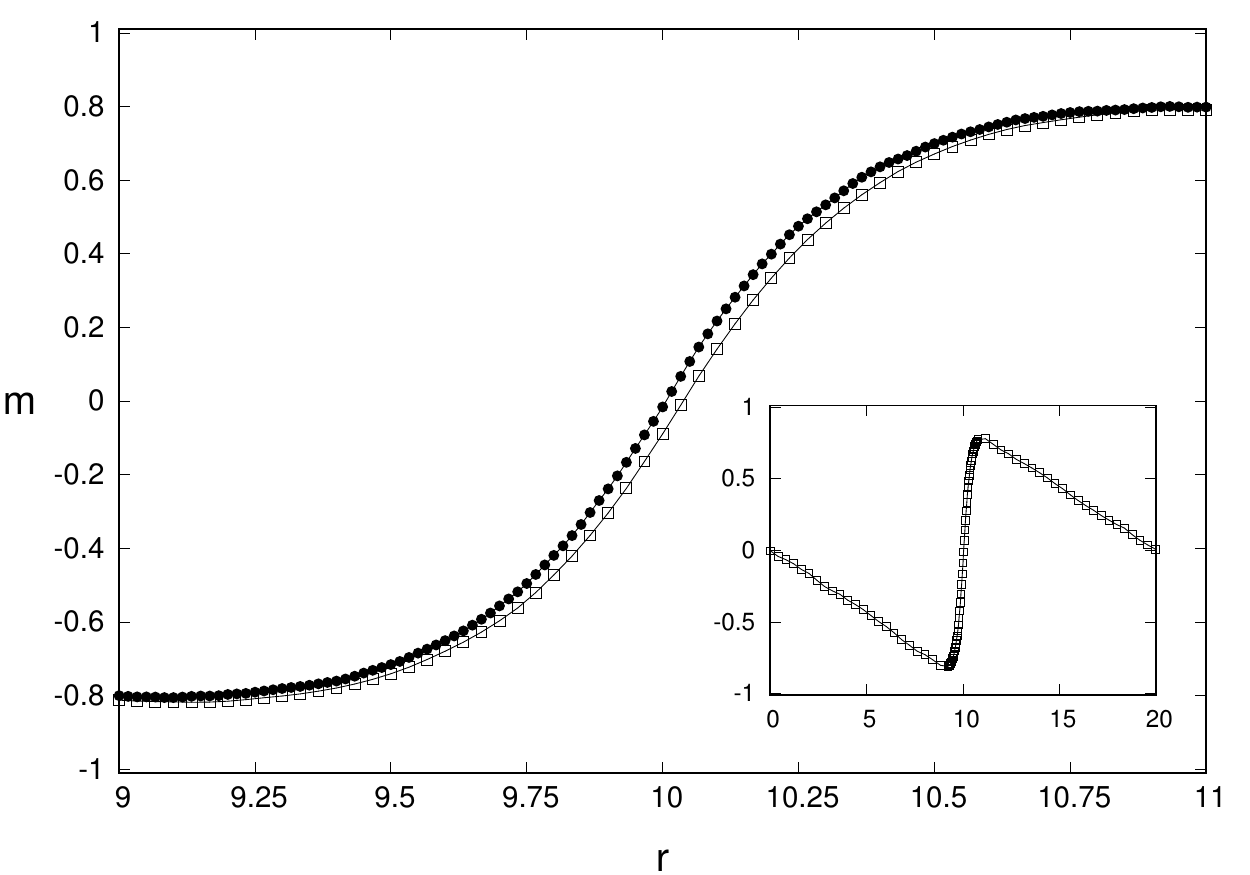}
\caption{Magnetization profile in mesoscopic units ($r=\ga x$)
in the presence of an external force $f_\ga(x)$, with $m_{+}=m_{-}=0$ and $\ell=20$. Shown is the comparison in $|r - \frac \ell 2| \le 1$ between the Monte Carlo prediction for $\gamma^{-1}=30$ and $L=600$ ($\smallsquare$) and for $\gamma^{-1}=60$ and $L=1200$ ($\smallblackcircle$). At the bottom right corner, the magnetization profile corresponding to $\gamma^{-1}=30$ and $L=600$ is shown over the whole interval $[0,\ell]$. }
\label{fig:fig2}
\end{figure}



\vskip1cm

The simulations show that the density profile is only weakly dependent on $\ga$, hence they suggest that it may have a limit when $\ga\to 0$.  Indeed, under suitable assumptions on the initial distribution and a propagation of chaos property we can prove, proceeding as in \cite{CDP_jsp}, that for any $t\ge 0$ and $r\in (0,\ell)$
  \begin{equation}
        \label{3.5}
\lim_{\ga \to 0}\;\;\lim_{\ga^2 t\to \tau; \ga x \to r}\;\;E[\si_t(x)] = m(r,\tau)
    \end{equation}
where $m(r,t)$ satisfies the conservation law
  \begin{equation}
        \label{3.6}
\frac {\partial} {\partial  t } m(r,t) = - \frac {\partial} {\partial  r } j(r,t),\quad m(0,t)= m_-,\;m(\ell,t)=m_+
    \end{equation}
The current $j=j(r,t)$ is equal to
 \begin{equation}
        \label{3.7}
j =\frac 12 \Big(-\frac {\partial m } {\partial  r }+
\beta (1-m^2 ) \{\la\int dx \rho_{\rm si}(x) \mathbf 1_{|r-x|\le 1} \big(
 \mathbf 1_{x < r}- \mathbf 1_{x \ge r}\big) \}\Big)
    \end{equation}
The curly bracket is the continuum version of the sum in \eqref{3.3}. In particular
  \begin{equation*}
j = -\frac 12 \frac {\partial m } {\partial  r }, \quad \text{for $|  r - \frac {\ell} 2|>1$ }
    \end{equation*}
To check the validity of  \eqref{3.6} we compare the profiles obtained in the simulations
with the stationary solution of  \eqref{3.6}:

\medskip
\noindent
{\em Stationary problem: find a constant $j^{(\ell)}$ and a function $m^{(\ell)}(r)$, $r\in [0,\ell]$, so
that $m^{(\ell)}(0)=m^{(\ell)}(\ell)=0$ and \eqref{3.7} is satisfied with
$j^{(\ell)}$ and $m^{(\ell)}$. }

\medskip
In Appendix \ref{appA} we prove:

\medskip

\begin{thm}
\label{thm3.1}
The above ``Stationary problem'' has for each $\ell$ a unique solution $\{j^{(\ell)}, m^{(\ell)}(x), x\in [0,\ell]\}$.  Such a solution has the following properties: $j^{(\ell)}>0$, $|m^{(\ell)}(x)| < 1$ for all $x$ and $\frac{d}{dx}m^{(\ell)}(x) = -2 j^{(\ell)}$ for all $|x - \frac{\ell}{2}| >1$.  Moreover
 \begin{eqnarray}
        \label{3.8}
&&\lim_{\ell\to\infty} m^{(\ell)}(\frac {\ell}{2}+r) = A(r),\quad |r|\le 1\\&& A(r) = \begin{cases}\tanh\{  {\beta( \la/2)}  [(1+r)^2-1]\}, & r\in[-1,0]\\
   \tanh\{   {\beta (\la/2)}  [1-(1-r)^2]\},& r\in[0,1] \end{cases} \nn
    \end{eqnarray}
while 
 \begin{equation}
        \label{3.9}
\lim_{\ell\to\infty} m^{(\ell)}(r\ell)  = M(r)=\begin{cases}- 2\alpha r, &  r\in [0,\frac 12)\\ 2\alpha(1-r), &  r\in (\frac 12 ,1]
\end{cases}, \qquad \alpha= A(1)=\tanh\{ {\beta \la/2} \}
    \end{equation}
Finally:
\begin{equation}
        \label{3.9.1}
\lim_{\ell\to\infty} \ell j^{(\ell)}  =  \tanh\{ {\beta \la/2} \}
    \end{equation}

\end{thm}

\begin{figure}
\centering
\includegraphics[width=0.6\textwidth]{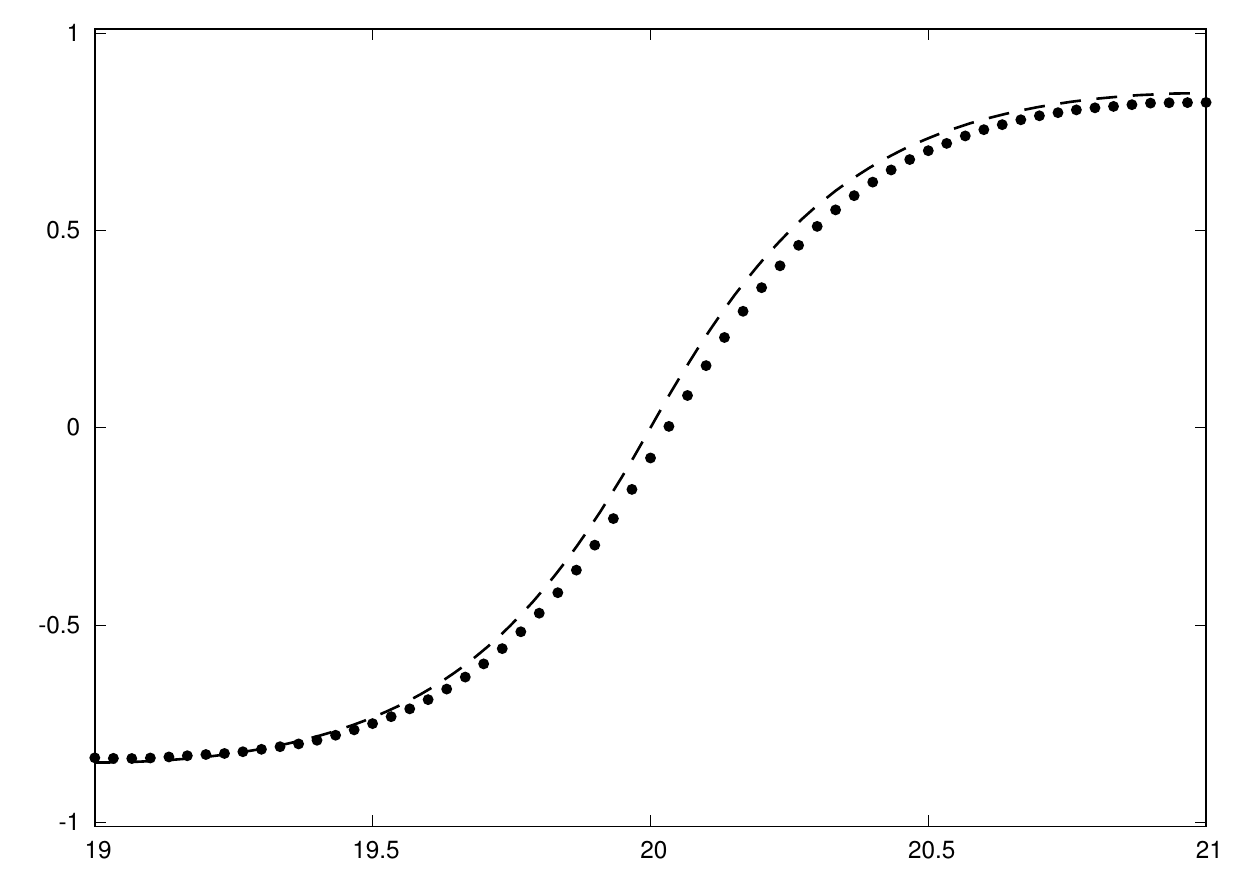}
\caption{Comparison between $\si^{t_0,T}(x)$ (black circles)  and $A(\ga x)$ (black dashed line)
in $|\ga x- \frac \ell 2| \le 1$ 
with $m_{+}=m_{-}=0$ and $\ell=40$, $\ga^{-1}=30$.}
\label{fig:fig3}
\end{figure}

\medskip
In \ref{fig:fig3} we compare the theoretical profile $A(r)$ and the average profile $\si^{t_0,T}(x)$ which indicates that we are with good approximation close to the mesoscopic and macroscopic limits when $\ga^{-1}=30$ and $L=1200$.
%
The measured current is $j_\pm ^{t_0,T}\approx 7.2 \cdot 10^{-4}$ so that by \eqref{3.9.1} and recalling that the mesoscopic current $j^{(\ell)}$ is related to the measured current by a factor $\ga^{-1}$ we get
\begin{equation}
        \label{3.10.1}
\left|j_\pm^{t_0,T}  -  \frac 1L\;\tanh\{ {\beta \la/2} \} \right| \le  2 \cdot 10^{-5}    
\end{equation}
which again shows that when $\ga^{-1}=30$ and $L=\ga^{-1}\ell=1200$ the system behaves with good approximation as in the macroscopic limit.


\vskip.5cm

\subsection{The mesoscopic theory}
The evolution equation \eqref{3.6} which describes the dynamics of our model in the limit $\ga\to 0$ has a nice physical interpretation.  In fact let $F(m)$,
 $m\in L^\infty([0,\ell],[-1,1])$, be the free energy functional
 \begin{equation}
        \label{3.11}
F(m) = \int dr \Big(- \frac {S(m(r))}{\beta}  +  m(r) U(r)
\Big)
    \end{equation}
where $S(m)$ is the entropy and $U(r)$ the potential generated by the silicon, namely
\begin{eqnarray}
\label{3.12}
 && S(m)= -\frac{1-m} {2}\log \frac{1-m} {2} - \frac{1+m} {2}\log \frac{1+m} {2}
 \\&& U(r) = \la \int_0^{\ell}dr'\rho_{\rm si}(r')J(|r-r'|)=\int_0^{\ell/2}dr'J(|r-r'|)
\end{eqnarray}
($J(|r|)$ being defined in \eqref{3.2}).
Then $j$ in \eqref{3.7} is equal to  
 \begin{equation}
        \label{3.13}
j = - \chi \frac{d}{dr}\frac{\delta}{\delta m(r)} F(m),\quad \chi = \frac{\beta}2 (1-m^2)
    \end{equation}
$\chi$ being the mobility (of the carbon atoms).  \eqref{3.13} is the usual constitutive law which states that the current is minus the mobility times the gradient of the chemical potential (which, according to thermodynamics, is the derivative of the free energy with respect to the density, recall that the density is $m(r)+1$).
In our case where there is a contribution to the free energy coming from the force exerted by the silicon, the current $j$ is not only given by minus the gradient of the density but it has an additional
contribution given by the second term in \eqref{3.7}. The curly bracket in  \eqref{3.7} has a clear physical meaning: the silicon atoms in the interval $dx$, i.e.\ $dx \rho_{\rm si}(x)$ generate a force field at $r$ which is $ \la\mathbf 1_{|r-x|\le 1}$ if $x<r$ (hence positive) and a negative force $ \la\mathbf 1_{|r-x|\le 1}$ if $x>r$.  The same expression can be rewritten as
 \begin{equation}
        \label{3.13.1}
\int dx U(|r-x|) \{\frac{d}{dx}\rho_{\rm si}(x)\}= U(|r- \frac \ell 2|)
   \end{equation}
Recalling that $U(|r-x|)=\la J(|r-x|)$ the left hand side of \eqref{3.13.1} is $\la$ times the weighted average of the density gradient of $\rho_{\rm si}(x)$. Thus the second term in \eqref{3.7} is $\la$ times the mobility times the averaged  density gradient of $\rho_{\rm si}$.  This is what expected from thermodynamics
if the averaging weight was a delta function.  Such an approximation would be valid if $m(r)$ were slowly varying at the edge, but this is not the case: when $\ell$ increases $m(r)$ becomes smoother but only away from the edge!

\begin{figure}
\centering
\includegraphics[width=0.6\textwidth]{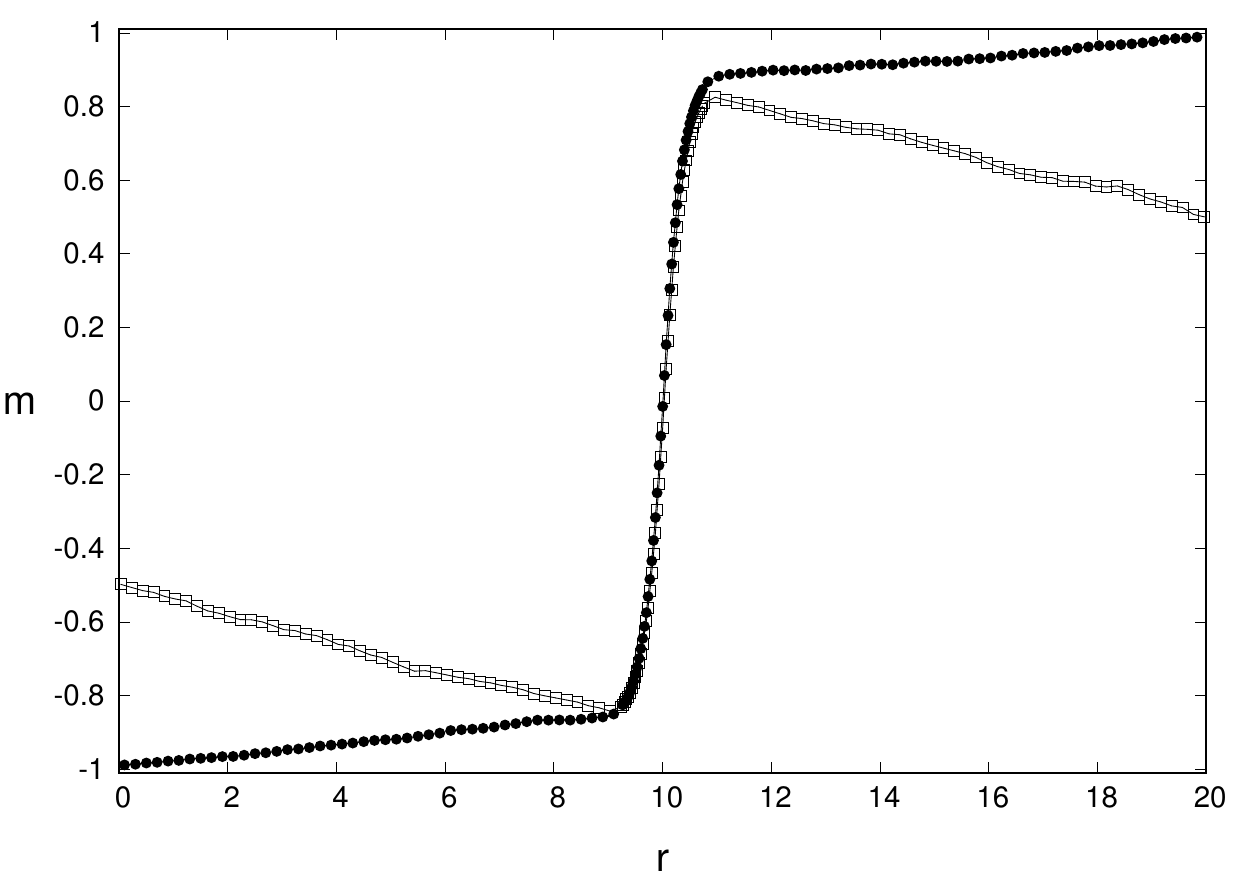}
\caption{Comparison between the Monte Carlo prediction for $m_{+}=0.5$ ($\smallsquare$) and $m_{+}=0.99$ ($\smallblackcircle$), with $\gamma^{-1}=30$, $L=600$, $m_{-}=-m_{+}$ and $\ell=20$. Note that, using $\beta=2.5$ and $\la=1$, we have $\tanh\{ {\beta \la/2} \}\approx 0.848$.}
\label{fig:fig4}
\end{figure}

The above proves that there is a
uphill diffusion with a mass flux from the left to the right reservoirs, despite they have the same density.  The work done to ensure such a flow is provided by the force exerted on carbon by the silicon atoms.  More generally we may take other densities for the reservoirs, for instance $m_+>0$ and
$m_-=-m_+$.  In such a case the analogue of Theorem \ref{thm3.1}   would give a positive current (and hence an uphill diffusion) if $m_+ < \tanh\{  {\beta \la/2} \}$, while the current would be downhill and hence negative if  $m_+ > \tanh\{ {\beta \la/2} \}$.
 see Fig. \ref{fig:fig4}

\vskip.5cm

\subsection{The macroscopic limit} The macroscopic limit is obtained by letting $\ell \to \infty$ while expressing the mesoscopic profiles in macroscopic units. By Theorem \ref{thm3.1} the limit profile is $M(r)$, $r\in [0,1]$, which by \eqref{3.9}  has a constant negative
slope $-2\tanh\{  \beta \la/2 \}$ except at $r= \frac 12$ where it jumps from $-  \tanh\{ \beta \la/2 \}$ to $\tanh\{ \beta \la/2 \}$.
Fick's law is satisfied and the differential equation in the macroscopic limit splits in two equations: one in $(0,\frac 12)$ and the other in  $(\frac 12,1)$.  The boundary conditions are $m(0)=0$   and $m(\frac 12) = -\tanh\{  \beta \la/2\} $ for the first equation and
$m(\frac 12) = \tanh\{ \beta \la/2\} $ and $m(1)=0$ for the second one.  Thus in the macroscopic scaling the action of the force (due to silicon) is represented by   boundary conditions at the discontinuity, this is a boundary layer problem which requires an analysis of the boundary layer in the stretched mesoscopic variables, as done in  Theorem \ref{thm3.1}.

 \bigskip

\setcounter{equation}{0}

\section{``Steady uphill diffusion'' and phase transitions}
\label{sec:4}

In this section we discuss
uphill diffusion due to phase separation referring to results obtained in \cite{CDP_pl},  \cite{CDP_jsp} and \cite{DPT}. As already mentioned in the introduction there are several surprising facts which are
not entirely understood.

The phase transitions
that we consider are of liquid-vapor type and the context is the
one proposed by van der Waals.  As explained by  van der Waals it is the presence of long range attractive forces which is responsible for the phase transition.  The way to implement his ideas in particle systems was first proposed by Kac with the introduction of Kac potentials, these are potentials which scale with a parameter $\ga>0$, the range scaling as $\ga^{-1}$ and the strength of the potential as $\ga^d$ ($d$ the dimension of the space), so that the total interaction of a point with the others stays finite as $\ga \to 0$. In \cite{KUH} and \cite{LP} it was shown that the Gibbsian statistical mechanics with Kac potentials reproduces the van der Waals theory in the limit $\ga \to 0$.

To implement all that we simply go back to the basic model of Section \ref{sec:2} and add an interaction of Kac type among particles.  This is just what we did in Section \ref{sec:3} but the force is now given by the same diffusing particles of the system and not by an external force (which in the previous section was exerted by the silicon atoms). In suitable units we suppose that the strength of the force (which in Section \ref{sec:3} was denoted by $\la$) is now equal to 1.
The model we obtain is that considered by the same authors in \cite{CDP_pl} and \cite{CDP_jsp}.
Referring to the system of Section \ref{sec:3} we only have to modify the velocity flip updating which is now as follows:

\vskip.5cm
\noindent
{\em Velocity flip}.  At all sites $x\in [1,L]$ where there is only
one particle we update its velocity  to become $+1$ with probability $\frac 12
+ \eps_{x,\ga}$ and $-1$ with probability $\frac 12
- \eps_{x,\ga}$, $\eps_{x,\ga}= C\ga^2[N_{+,x,\ga}-N_{-,x,\ga}]$; at all other sites the occupation numbers are left unchanged.  We have set
\begin{equation}
\label{4.1}
N_{+,x,\ga}
= \sum_{y=x+1}^{ x+\ga^{-1}}\eta^{(+)}(y),\;
N_{-,x,\ga}
= \sum_{y= x-\ga^{-1}}^{x-1}\eta^{(-)}(y),\quad x \in [1,L]
\end{equation}
where $\eta^{(+)}(y)= \eta(y)$ if $y \in [1,L]$
and $\eta^{(+)}(y)= 2\rho_{+}$ if $y >L$; similarly  $\eta^{(-)}(y)= \eta(y)$ if $y \in [1,L]$
and $\eta^{(-)}(y)= 2\rho_{-}$ if $y <1$, recall that $2\rho_{\pm}$ is the density of the right, respectively left reservoir.
We choose
$C= 1.25$ and $\ga^{-1}= 30$ so that the definition is well posed because $(2\ga^{-1}) C\ga^2= 2.5/30<\frac 12$, $(2\ga^{-1}) $ being an upper bound for $|N_{+,x,\ga}-N_{-,x,\ga}|$.

\vskip.5cm

By \eqref{3.4} and recalling that $\la=1$ the above choice implies that particles are in contact with an environment which keeps the inverse temperature $\beta$ equal to
\begin{equation}
\label{4.2}
\beta=2C= 2.5
\end{equation}
In Section 5 of \cite{CDP_jsp} it is shown that in the limit $\ga\to 0$ considered in the previous section the evolution is ruled again by the conservation law \eqref{3.6} with the current $j(r,t)$ given by \eqref{3.13} where the free energy functional $F(m)$ is now given by
\begin{eqnarray}
\label{4.3}
 &&F(m) = \int \Big(-\frac {m^2}{2} - \frac {S(m)}{\beta}\Big) + \frac 14\int\int
 J (r,r') [m(r)-m(r')]^2
\end{eqnarray}
with $m(r)=m_{\pm}$ if $r \ge \ell$ and respectively $r \le 0$.
The first term on the right hand side, namely
\begin{eqnarray}
\label{4.4}
 &&
 f_\beta(m) := -\frac {m^2}{2} - \frac {S(m)}{\beta}
\end{eqnarray}
is the van der Waals mean field free energy, which is a convex function for $\beta\le 1$ while for $\beta>1$ becomes a double well with minima at $\pm m_\beta$ where
\begin{eqnarray}
\label{4.5}
 &&
  m_\beta =  \tanh\{\beta m_\beta\}, \quad m_\beta >0
\end{eqnarray}
In our model $\beta = 2.5$ hence we are in the phase transition regime.
The values $|m| \ge m_\beta$ define the stable phases, the interval $|m| < m_\beta$ is the spinodal region.  Inside the spinodal region the set $|m| < m^*$, $m^*>0 :  \beta(1-(m^*)^2) =1$, is  unstable  while the region  $m^* <|m| < m_\beta$ is metastable.

The current $j(r,t)$ given by \eqref{3.13} with $F(m)$ as in \eqref{4.3} is:
 \begin{equation}
\label{4.6}
 j(r,t)= -\frac 12\big\{\frac{\partial m(r,t)}{\partial r}
 - \beta[1-
 m(r,t)^2]\int dx J(|r-x|) \frac{\partial m(x,t)}{\partial x}
  \big\}
 \end{equation}
If we suppose that $\frac{\partial m(x,t)}{\partial x}\approx
\frac{\partial m(r,t)}{\partial r}$ in the support of $J(|r-x|)$ then
 \begin{equation}
\label{4.7}
 j(r,t)\approx -\frac 12\frac{\partial m(r,t)}{\partial r}\big\{1
 - \beta[1-
 m(r,t)^2]
  \big\}
 \end{equation}
Thus when $|m(r,t)| < m^*$, i.e.\ in the unstable region, the current has the same sign of the gradient of $m$ and the diffusion coefficient is negative.  This is at the basis of the macroscopic explanation of the uphill diffusion.

In our context the approximation $\frac{\partial m(x,t)}{\partial x}\approx
\frac{\partial m(r,t)}{\partial r}$ in the support of $J(|r-x|)$ is shaky because in the spatial region where $m$ is unstable its values vary significantly.
In our simulations  we set $m_+=-m_->0$ and decrease $m_+$ starting from its maximal value 1.

\begin{figure}
\centering
\includegraphics[width=0.6\textwidth]{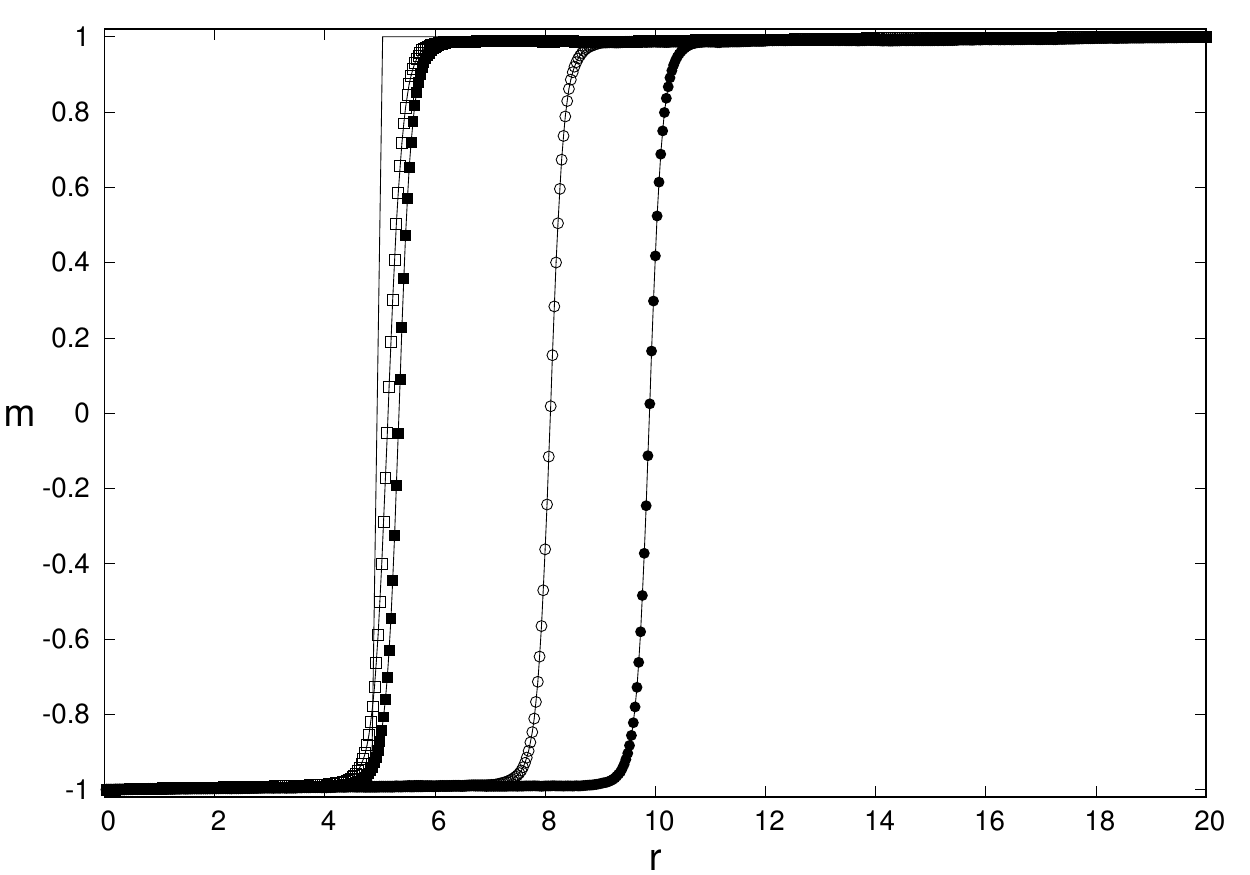}
\caption{Magnetization profile in presence of a Kac potential among particles, for $C=1.25$
and  $m_+ =1$, with space in $\ga^{-1}$ ($=30$) units. The parameters $m_\beta$ and $m^{*}$ have values $m_\beta=0.985$ and $m^{*}=0.775$.  The different curves in the plot correspond to the averaged magnetization computed at different times: $t_0=10^5$ ($\smallsquare$), $t_0=10^6$ ($\smallblacksquare$), $t_0=10^7$ ($\smallcircle$) and $t_0=10^8$ ($\smallblackcircle$). The black thin line denotes the initial configuration, given by a step function centered at $r=15$.}
\label{fig:fig5}
\end{figure}

\begin{figure}
\centering
\includegraphics[width=0.6\textwidth]{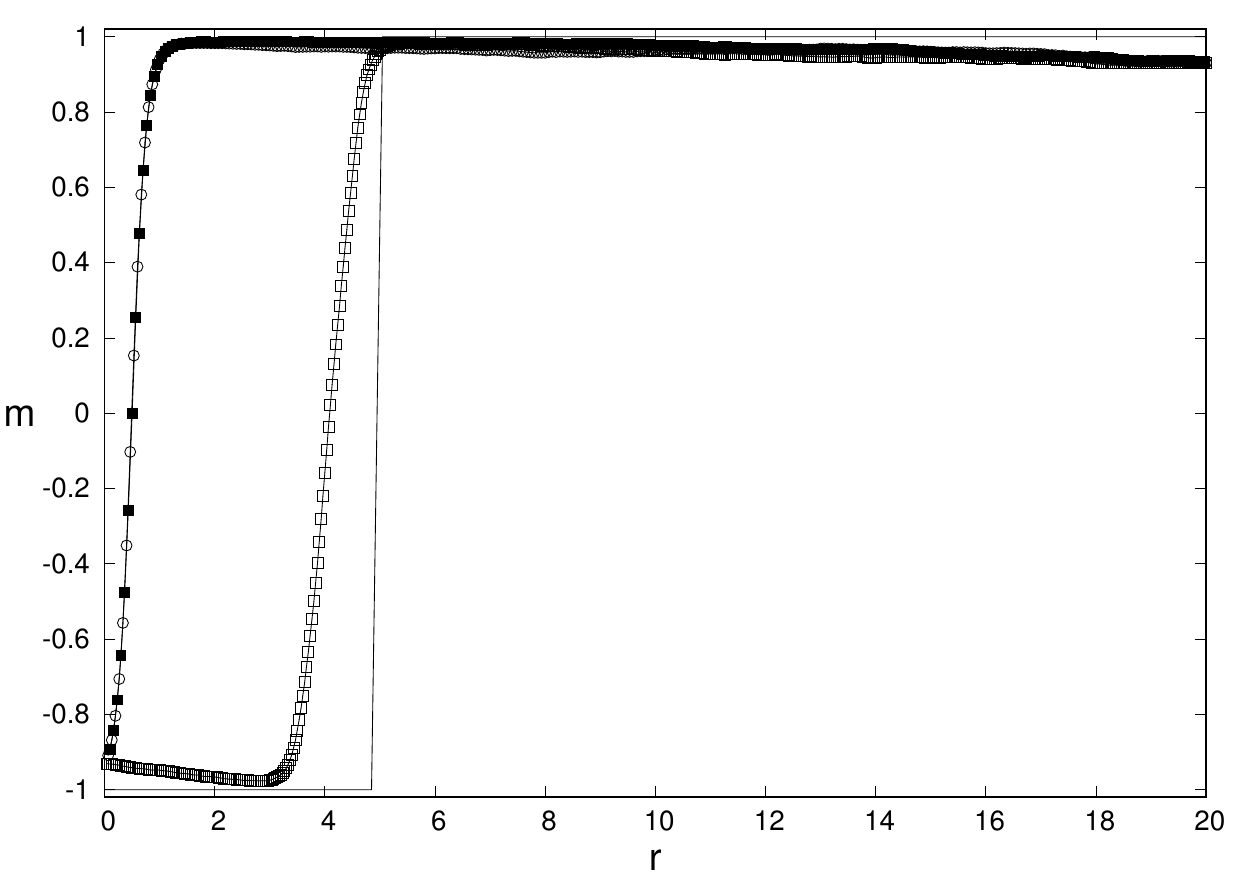}
\caption{Magnetization profile with $m_{+}=0.93$. The different curves in the plot correspond to the averaged magnetization computed at different times: $t=10^5$ ($\smallsquare$), $t=10^6$ ($\smallblacksquare$) and $t=10^8$ ($\smallcircle$). The black thin line denotes the initial configuration, corresponding to a step function centered at $r=5$}
\label{fig:fig6}
\end{figure}

In Fig. \ref{fig:fig5} we see a negative (downhill) current till $m_+ > m_\beta$ while in Fig. \ref{fig:fig6} it becomes positive as $m_+< m_\beta$, in this latter case the current goes uphill from the reservoir with smaller to the one with larger density. The stationary profile has also a significant change, when $m_+> m_\beta$ it is smooth away from a small neighborhood of the middle point, where instead has a sharp jump going from $\approx -m_\beta$ to $m_\beta$.  Instead when $m_+<m_\beta$ and metastable (i.e.\ $m_+ > m^*$)
the jump moves to one of the endpoints, in the simulation presented in Fig. \ref{fig:fig6} it goes to the left boundary where it jumps from $m_-$ to a value $b(m_-)$ which is larger than $m_+$, it then decreases smoothly toward the value $m_+$ reached at the right boundary.

When $m_+ < m^*$ the current is still positive  (i.e.\ uphill) but the profile has a more complex structure, we refer to \cite{CDP_jsp} for details. In \cite{CDP_jsp} we give some theoretical explanation of these phenomena but a complete theory with mathematical proofs is still missing.

The canonical system in statistical mechanics to study phase transitions is the Ising model in $d \ge 2$ dimensions with ferromagnetic nearest neighbor interactions.  There are preliminary
results with computer simulations in the $d=2$ case (obtained by Colangeli, Giardin\`a, Giberti, Vernia)  which show again uphill diffusion in essential agreement with what described above.

\bigskip
\noindent
{\bf Acknowledgments:}

M.C. acknowledges useful discussions with Ellen Moons and Andrea Muntean (Karlstad University).

\bigskip
\appendix

\section{The stationary problem}
\label{appA}

In this section we prove Theorem \ref{thm3.1}.
We will first  prove in Corollary \ref{corollaryA.2} below the existence of
stationary solutions, namely we will prove that
there are a constant $j$ and a function $m(x)$, $x \in [0,\ell]$, such that
 \begin{eqnarray}
 \label{A.1}
 \frac {d m } {d x }=-2j + \beta \la
 [1-m^2(x) ]g(x),\qquad m(0)= m(\ell)=0
    \end{eqnarray}
    where
        \begin{eqnarray}
    \label{A.2}
   g(x)&=& \int_{x-1}^{x}  \mathbf 1_{y\le \frac \ell 2}\,\,dy -\int_{x }^{x+1}  \mathbf 1_{y\le \frac \ell 2}\,\,dy
    \end{eqnarray}
    In particular
  \begin{equation}
        \label{A.3}
 \frac {d m } {d  x }=-2 j, \quad \text{for $|  x - \frac {\ell} 2|>1$ }
    \end{equation}
Observe that    $ h(x):=g(\frac \ell 2 +x)$, $|x| \le 1$, is equal to
    \begin{eqnarray}
    \label{A.2.1}
   h(x) &=&\mathbf 1_{x\in [-1,0]}(1+x)+ \mathbf 1_{x\in (0,1]}(1-x)
    \end{eqnarray}
and therefore independent of $\ell$.

We start by proving the following lemma:

\medskip

   \begin{lemma}
   \label{lemmaA.1}
Denote by $m_{(j)}(x)$, $x \ge 0$, $j\in\mathbb R$, the solution of
\begin{eqnarray}
      \label{A.4}
 \frac {d m } {d  x }=-2j +
\beta \la [1-m^2 ]g(x),\qquad m(0)=0
    \end{eqnarray}
 Then
 	\begin{eqnarray}
      \label{A.5}
      m_{(j')}(\ell)< m_{(j)}(\ell),\quad \text{if}\quad  j'> j,\qquad \lim_{j\to\pm \infty}m_{(j)}(\ell)=\mp \infty
          \end{eqnarray}
\end{lemma}

\noindent{\bf Proof.}
Take $j'> j$. Then by \eqref{A.3}  $m_{(j)}(x)=-2jx> m_{(j')}(x)=-2j'x$ for $x\in (0,\frac \ell 2-1]$. Suppose by contradiction that there is  $y> \frac \ell 2-1$ such that
$m_{(j)}(x)>m_{(j')}(x)$ for $x<y$ and $m_{(j)}(y)=m_{(j')}(y)$. Then $(1-m_{(j)}(y)^2 )g(y)=(1-m_{(j')}(y)^2 )g(y)$ and therefore
   $$
   \frac {d}{dy} \Big(m_{(j)}(y)-m_{(j')}(y)\Big)  =2(j'-j)>0
   $$
which contradicts the inequality $m_{(j)}(x)>m_{(j')}(x)$ valid for $x<y$,
hence the first statement in \eqref{A.5}.

To prove the second  statement in \eqref{A.5} we first consider $j<0$.  In such a case
$ \dis{\frac {d  } {d x  }m_{(j)}(x) \ge -2 j >0}$ for all $x$, hence for any $x>0$
$\lim_{j\to -\infty}m_{(j)}(x) = \infty$.  When $j>0$ we define
  \begin{equation}
        \label{A.6}
 \mathcal X:=\{x>0:
m_{(j)}(x) \le -1\}
    \end{equation}
In $ \mathcal X $ we have $ \dis{\frac {d  } {d x  }m_{(j)}(x) \le -2 j <0}$. As a consequence
if
$ y\in \mathcal X$ then $ x\in \mathcal X$  for all $x \ge y$.  For $j$ large enough
$\frac \ell 2 -1 \in  \mathcal X$ and therefore
     $$
 m_{(j)}(x) \le -1 - 2j (x-(\frac \ell 2 -1)), \quad x > \frac \ell 2 -1
     $$
hence $\lim_{j\to\infty}m_{(j)}(\ell) = -\infty$.
\qed

\medskip

   \begin{corollary}
   \label{corollaryA.2}
There is a unique solution $\{j^{(\ell)},m^{(\ell)}(x), x\in [0,\ell]\}$   of \eqref{A.1}. Furthermore $j^{(\ell)}>0$
 and
 $|m^{(\ell)}(x)| < 1$ for all $x \in [0,\ell]$.
    \end{corollary}

\noindent{\bf Proof.} $m_{(j)}(\ell)$ is a continuous, strictly decreasing function of $j$ which converges to $\pm \infty$ as $j \to \mp \infty$, hence there is a unique $j^{(\ell)}$ such that
$m_{(j^{(\ell)})}(\ell)=0$ and $m^{(\ell)}(x) := m_{(j^(\ell))}(x)$, hence it is the unique solution of \eqref{A.1}.

Call for notational simplicity $m(x)$ and $j$ the  unique solution
of \eqref{A.1}.  Suppose by contradiction that $j \le 0$, then, by \eqref{A.3}, $dm(x)/dx \ge 0$ for all $x$ and
$dm(x)/dx > 0$ for some $x \in (\frac \ell 2 -1,\frac \ell 2 +1)$, which yields $m(\ell)>0$, while $m(\ell)=0$ by \eqref{A.1}, thus $j>0$.

Suppose again by contradiction that there is $y$ such that $m(x) < 1$ for all $x<y$ and
$m(y)=1$. By \eqref{A.3}, $dm(y)/dy=-2j <0$ hence the contradiction because it would mean that $m(x)>1$ for $x<y$ and $y-x$ small enough.  Suppose again by contradiction that there is $y < \ell$ such that $m(y)=-1$, then $y \in \mathcal X$, see \eqref{A.6}, and as argued in the proof of Lemma \ref{lemmaA.1}, this implies $\ell \in \mathcal X$, while $m(\ell)=0$ by \eqref{A.1}.
  \qed

\medskip

   \begin{lemma}
   \label{lemmaA.3}
   Let $\{j^{(\ell)},m^{(\ell)}\}$ be as in Corollary \ref{corollaryA.2}, then
    \begin{eqnarray}
      \label{A.7}
 2j^{(\ell)} = \frac{- m^{(\ell)}(\frac \ell 2 -1)}{\frac \ell 2 -1}
  < (\frac \ell 2 -1)^{-1},\quad m^{(\ell)}(\frac \ell 2 +1)=-
 m^{(\ell)}(\frac \ell 2 -1) >0
    \end{eqnarray}

\end{lemma}

\noindent
{\bf Proof.}  By \eqref{A.3} $m^{(\ell)}(\frac \ell 2 -1)= -2j^{(\ell)}(\frac \ell 2 -1)$
and $m^{(\ell)}(\frac \ell 2 +1)=  j^{(\ell)}(\ell-[\frac \ell 2 +1])$ hence \eqref{A.7} having used that $|m^{(\ell)}| < 1$.  \qed

\vskip.5cm

Let $a(x|\alpha)$, $|x| \le 1$, $\alpha \in (-1,0)$, be the solution of
 \begin{eqnarray}
 \label{A.8}
 \frac {d a } {d x }= \beta \la
 [1-a^2(x) ]h(x),\quad a(-1)= \alpha
    \end{eqnarray}
where $h(x)$ has been defined in \eqref{A.2.1}.  Explicitly:
 \begin{equation}
 \label{A.9}
 a (x |\alpha)= \begin{cases}\tanh\big\{\beta (\la/2) (1+x)^2 + \tanh^{-1}(\alpha)\big\}, & x \le 0\\
 \tanh\big\{\beta (\la/2) [1-(1-x)^2] + \tanh^{-1}(a (0 |\alpha))\big\}, &x > 0
 \end{cases}
    \end{equation}

\medskip

   \begin{lemma}
   \label{lemmaA.4}
   Let $a^{(\ell)}(x)=a(x|\alpha)$  with
   $\alpha=m^{(\ell)}(\frac \ell 2 -1)$.
   Then
    \begin{eqnarray}
      \label{A.10}
 \sup_{|x| \le 1}|m^{(\ell)}(\frac \ell 2 +x) -a^{(\ell)}(x)| \le  \frac{j^{(\ell)}}{2\beta \la}
 e^{4\beta \la}
    \end{eqnarray}

\end{lemma}

\noindent
{\bf Proof.} Since $|m^{(\ell)}| <1$ and, by \eqref{A.9}, $|a^{(\ell)}|<1$
   \begin{eqnarray}
      \label{A.11}
  |\frac{d}{dx}\Big(m^{(\ell)}(\frac \ell 2 +x) -a^{(\ell)}(x)\Big)| \le j^{(\ell)}
  + \beta \la h(x) |m^{(\ell)}(\frac \ell 2 +x)- a^{(\ell)}(x)| 2
    \end{eqnarray}
hence \eqref{A.10}.  \qed

\bigskip

\noindent
{\bf Proof of \eqref{3.8}.}  Let $\ell_n$ be any sequence such that
$\ell_n \to \infty$ and
such that $m^{(\ell_n)}(\frac {\ell_n} 2 -1)$ has a limit, call it $-\alpha^*$.  Then
by  \eqref{A.7} $\alpha^* >0$ and $m^{(\ell_n)}(\frac {\ell_n} 2 +1)\to \alpha^*$.  By Lemma \ref{lemmaA.4} and \eqref{A.9}
  $$
  \lim_{n\to \infty} m^{(\ell_n)}(\frac {\ell_n} 2 +x) =\lim_{n\to \infty}a^{(\ell_n)}(x)= a(x|-\alpha^*)
  $$
and $a(1|-\alpha^*)=\alpha^*$. By \eqref{A.9}
	\begin{eqnarray*}
\alpha^*=a(1|-\alpha^*)=\tanh\big \{\beta\la -\tanh^{-1}\alpha^*\big\}
	\end{eqnarray*}
and this implies $\alpha^* = \tanh\{\beta \la\}$. \qed

\bigskip

\noindent
{\bf Proof of  \eqref{3.9.1}.} From  \eqref{A.7} and  \eqref{3.8} we have
$$
\ell j^{(\ell)} = -\frac{\ell }{ \ell -2} m^{(\ell)}(\frac \ell 2 -1)=\frac{\ell }{ \ell -2} m^{(\ell)}(\frac \ell 2 +1)\to \alpha^*=\tanh\{\beta\la/2\}
$$
\qed

\bigskip

\noindent
{\bf Proof of  \eqref{3.9}.}  Let $r\in [0,\frac 12)$ then
 \begin{equation*}
 m^{(\ell)}(r\ell)  = - 2 j^{(\ell)}r\ell \to -2\alpha r
 \end{equation*}
The analogous statement holds for $r > 1/2$ and \eqref{3.9} is proved.
 \qed


\begin{thebibliography}{3}

\bibitem{Alvarez} J. Alvarez-Ramirez, L. Dagdug, M. Meraz, {\em Asymmetric diffusion in heterogeneous media}, Physica A {\bf 395}, 193--199 (2014).

\bibitem{CDP_pl}  M. Colangeli, A. De Masi, E. Presutti, {\em{ Latent heat and the Fourier law}}, Phys. Lett. A \textbf{380}, 1710--1713 (2016).

\bibitem{CDP_jsp} M. Colangeli, A. De Masi, E. Presutti, {\em{Particle models with self sustained current}} J. Stat. Phys. (2017);  doi:10.1007/s10955-017-1765-3.


\bibitem{darken}  L. S. Darken, {\em{Diffusion of carbon in austenite with a discontinuity in composition}}, Trans. AIME {\bf 180}, 430 (1949).

\bibitem{DPT}
A. De Masi, E. Presutti, D. Tsagkarogiannis, {\em  Fourier law, phase transitions and the stationary Stefan problem } Arch. Rational Mech. Anal. \textbf{201}, 681--725 (2011).

\bibitem{Diebel} M. Diebel, S. T. Dunham. {\em Ab Initio Calculations to Model Anomalous Fluorine Behavior}, Phys. Rev. Lett. {\bf 93}, 245901 (2004).

\bibitem{Erleb} J. Erlebacher, M. J. Aziz, A. Karma, N. Dimitrov, K. Sieradzk, {\em Evolution of nanoporosity in dealloying}, Nature {\bf 410}, 450--453 (2001).

\bibitem{Frink} L. J. D. Frink, A. Thompson, A. G. Salinger, {\em Applying molecular theory to steady-state diffusing systems},  J. Chem. Phys. {\bf 112}, 7564 (2000).

\bibitem{KUH} M. Kac, G. E. Uhlenbeck, P. C. Hemmer, {\em On the van der Waals theory of vapor-liquid equilibrium.I. Discussion of a one dimensional model}. J. Math. Phys. {\bf 4}, 216 (1963);
{\em On the van der Waals theory of vapor-liquid equilibrium.
II. Discussion of the distribution functions.} J. Math. Phys. {\bf 4}, 229 (1963);
{\em  On the van der Waals theory of vapor-liquid equilibrium.
III. Discussion of the critical region.} J. Math. Phys. {\bf 5}, 6 (1964).

\bibitem{Karg} J. K\"{a}rger, D. M. Ruthven, {\em Diffusion in nanoporous materials: fundamental principles, insights and challenges}, New J. Chem. {\bf 40}, 4027--4048 (2016).

\bibitem{RK}  R. Krishna, {\em Uphill diffusion in multicomponent mixtures}
Chem. Soc. Rev., {\bf 44}, 2812--2836 (2015).

\bibitem{Lauer} A. Lauerer, T. Binder, C. Chmelik, E. Miersemann, J. Haase, D. M. Ruthven, J. K\"{a}rger, {\em Uphill diffusion and overshooting in the adsorption of binary mixtures in nanoporous solids}, Nat. Commun. {\bf 6}, 7697 (2015).

\bibitem{LP}  J.L. Lebowitz, O. Penrose, {\em Rigorous Treatment of
the Van Der Waals-Maxwell Theory of the Liquid-Vapor Transition.}
J. Math. Phys. {\bf 7}, 98--113  (1966).

\bibitem{Sato} N. Sato, Z. Yoshida, {\em Up-hill diffusion, creation of density gradients: Entropy measure for systems with topological constraints}, Phys. Rev. E {\bf 93}, 062140 (2016).

\bibitem{Tsuc} S. Tsuchiya, M. Seno, {\em Uphill Transport of Metal Picrate through a Liquid Membrane by Ylides}, J. Phys. Chem. {\bf 98}, 13680--13686 (1994). 

\bibitem{Vitag} V. Vitagliano, R. Sartorio, S. Scala, D. Spaduzzi, {\em Diffusion in a ternary system and the critical mixing point}, J. Solution Chem. {\bf 7}, 605--622 (1978). 

\end{thebibliography}
 \end{document}